\documentclass{webofc}
\usepackage[varg]{txfonts}   
\include{amsmath}
\include{amssymb}
\begin{document}
\title{Semileptonic $\tau$ decays: powerful probes of non-standard charged current weak interactions}
%
%

\author{\firstname{Pablo} \lastname{Roig}\inst{1}\fnsep\thanks{\email{proig@fis.cinvestav.mx}}}

\institute{Centro de Investigaci\'on y de Estudios Avanzados del IPN, Apdo. Postal 14-740, 07000 Ciudad de M\'exico, M\'exico.}

\abstract{%
  When looking for heavy ($\mathcal{O}$(few TeV)) New Physics, the most efficient way to benefit from both high 
  and low-energy measurements simultaneously is the use of the Standard Model Effective Field Theory (SMEFT). In this talk I highlight the importance of semileptonic $\tau$ decays in complementing, in this respect, the traditional low-energy precision observables 
  and high-energy measurements. This is yet another reason for considering hadronic tau decays as golden channels at Belle-II beyond the unquestionable interest of the CP violation anomaly in $\tau\to K_S\pi\nu_\tau$ decays, that I also discuss within the effective theory. A couple of new results for $\tau^-\to K^-\nu_\tau$ decays are also included.
}
\maketitle
\section{Introduction: Effective field theory application}
\label{intro}
I explain in this section the first sentence of the abstract above.

Imagine four fermion weak interactions are modified, with respect to the Standard Model (SM) \footnote{Where they can proceed, e. g. through $W$ exchange at tree level.}, by the exchange of new heavy mediators with $\mathcal{O}(10)$ TeV masses. At this energy scale one must include all renormalizable operators of dimension $4$ consistent with the underlying local gauge symmetry in the Lagrangian, including both SM and beyond the SM (BSM) degrees of freedom. In principle this could be \textit{the fundamental theory} but, at least, it will be an extension of the SM with an increased energy range of validity, reaching length scales smaller than probed experimentally so far. If, instead, we look at the same processes at $E\sim 1$ TeV, then the heavy BSM degrees of freedom are no longer dynamical and they should be integrated out from the action. This will give rise (in addition to the SM Lagrangian) to an infinite tower of (classically) non-renormalizable operators written in terms of the SM fields. The higher their dimension, the larger their suppression, given by inverse powers of the new physics (NP) scale. This is the SMEFT Lagrangian \cite{Buchmuller:1985jz, Grzadkowski:2010es} \footnote{In this setting, it is assumed that the NP is weakly coupled at few TeV and linear realization of the spontaneous electroweak symmetry breaking.} and, given the absence of NP detected at the LHC \footnote{Limits on specific realizations are in the few TeV range.}, it is the most effective way of analyzing LHC data in a model-independent way. For incorporating the electroweak precision observables (EWPO) one needs to run down the theory up to the $M_Z$ scale (with the EW$\times$QCD gauge group) \cite{Alonso:2013hga}. Still, if we want to benefit also from the very high-precision data obtained at low energies, we have to run again down (with the electromagnetic times strong gauge group) up to the typical mass scale of the problem \cite{Gonzalez-Alonso:2017iyc} ($\mu_{low}:=2$ GeV$\sim M_\tau$ in our case) ensuring the proper matching at $M_Z$. At these low energies, all SM heavy (compared to $\mu_{low}$) fields also need to be integrated out. As a result, in the low-energy limit of SMEFT, the leading contributions to the considered processes come from dimension six operators that give rise to a Fermi-type local interaction. The main advantage of the EFT framework glows: different experimental measurements and their implications for NP can be compared unambiguously in a model-independent way either at low energies or at LEP or LHC scales.

In our cases of interest the corresponding Lagrangian can be written as
\cite{Cirigliano_Jenkins_Gonzalez,Bhattacharya_Cirigliano},
\begin{equation}
    \begin{split}
            \mathcal{L}&=-\frac{G_{F}V_{uD}}{\sqrt{2}}(1+\epsilon_{L}+\epsilon_{R})\Big[\bar{\tau}\gamma_{\mu}(1-\gamma_{5})\nu_{\ell}\cdot\bar{u}[\gamma^{\mu}-(1-2\hat{\epsilon}_{R})\gamma^{\mu}\gamma_{5}]D\\
        &+\bar{\tau}(1-\gamma_{5})\nu_{\ell}\cdot\bar{u}[\hat{\epsilon}_{s}-\hat{\epsilon}_{p}\gamma_{5}]D+2\hat{\epsilon}_{T}\bar{\tau}\sigma_{\mu\nu}(1-\gamma_{5})\nu_{\ell}\cdot\bar{u}\sigma^{\mu\nu}D\Big]+\mathrm{h.c.}\,, \label{eqL}
        \end{split}
\end{equation}
with $D=d,s$ (an upper-index will be used when distinguishing $\epsilon$ couplings involving different lepton families but the $D$-type quark involved is to be understood). Wrong-flavor or -helicity neutrinos do not contribute at linear order in the NP effective couplings (in which $\hat{\epsilon}_i:=\epsilon_i/(1+\epsilon_{L}+\epsilon_{R})\sim \epsilon_i$) and thus have been neglected.

Assuming natural values of the NP effective couplings at a few TeV, the effective NP energy scale probed  is $ \Lambda\sim v(V_{uD}\epsilon_i)^{-1/2}$ \footnote{The $\epsilon$ couplings are related to the relevant SMEFT couplings in appendix A of Ref.~\cite{Bhattacharya_Cirigliano}.}.

The main results of applying eq. (\ref{eqL}) to hadronic tau decays in searching for heavy NP in a model-independent way will be discussed, and compared to other low- and high-energy probes in the remainder of this contribution.
\section{Application to semileptonic tau decays}
\label{sec-App}
We will start discussing the determinations from exclusive decay channels and consider later on those from inclusive analysis for the strangeness conserving and changing cases, in turn.
\subsection{$\tau^-\to\pi^-\nu_\tau$}
\label{sec-pi}
Since this process is mediated by the axial-vector current, it is sensitive to $|\epsilon_L^\tau-\epsilon_R^\tau|$. In addition to the SM pseudoscalar piece induced by the partial conservation of the axial-vector current one could in principle have \textit{genuine} pseudoscalar contributions encoded in $\epsilon_P^\tau$. Finally, since the extraction of $V_{ud}$ from superallowed Fermi transitions may hide NP contributions in the vector current ($\sim\epsilon_L^e+\epsilon_R^e$), the $\tau^-\to\pi^-\nu_\tau$ decays determine \cite{Cirigliano:2018dyk}
\begin{equation}\label{eq1pi}
 \epsilon_L^\tau-\epsilon_L^e-\epsilon_R^\tau-\epsilon_R^e-\frac{m_\pi^2}{M_\tau(m_u+m_d)}\epsilon^\tau_P\,=\,(-1.5\pm6.7)\times 10^{-3}\,
\end{equation}
at the subpercent level (provided radiative corrections are included). Again, one cannot use the pion decay constant determined from data (as it may include NP effects) but rather employ its lattice QCD determination \cite{Tanabashi:2018oca}. This is the current main source of uncertainty in the previous result. The potential of combining different data sets is illustrated by taking the ratio with the $\pi$ decay to cancel the dependence on $F_\pi$ (and $V_{ud}$), yielding
\cite{Cirigliano:2018dyk}
\begin{equation}\label{eqratiowithpi}
 \epsilon_L^\tau-\epsilon_L^\mu-\epsilon_R^\tau+\epsilon_R^\mu-\frac{m_\pi^2}{M_\tau(m_u+m_d)}\epsilon^\tau_P+\frac{m_\pi^2}{m_\mu(m_u+m_d)}\epsilon^\mu_P\,=\,(-3.8\pm2.7)\times 10^{-3}\,
\end{equation}
with an uncertainty reduced more than twice.
\subsection{$\tau^-\to\pi^0\pi^-\nu_\tau$}
\label{sec-pipi}

In this case one needs to rely on theory, and two strategies have been proposed. These will be examined in turn next. One trusts:\\
\textbf{a)} The computed isospin-breaking corrections relating di-pion $\tau$ decays and $e^+e^-$ data \cite{Davier:2009ag}.\\
\textbf{b)} Dispersive representations of the participant form factors fitted to data.\\

Concerning \textbf{a)}, the key feature is exploiting that any possible heavy NP effect in the neutral current will be negligible compared to the SM photon exchange. Thus, a difference between results obtained with charged weak and neutral electromagnetic currents can be interpreted as a NP contribution to the tau decays, in such a way that
\begin{equation}\label{eqamu}
 \frac{a_\mu^\tau-a_\mu^{ee}}{2a_\mu^{ee}}\,=\,\epsilon^\tau_L-\epsilon^e_L+\epsilon^\tau_R-\epsilon^e_R+1.7\epsilon^\tau_T\,=\,(8.9\pm4.4)\times10^{-3}\,.
\end{equation}

As in eq.~(\ref{eq1pi}), the dependence on $\epsilon^e_L+\epsilon^e_R$ comes from the determination of $V_{ud}$. Eq.~(\ref{eqamu}) rephrases the well-known \cite{Davier:2009ag} difference between the two-pion contributions to the hadronic vacuum polarization part of the anomalous magnetic moment of the muon obtained with either $e^+e^-$ or $\tau$ data as a $\sim2\sigma$ hint for BSM contributions to the relevant combination of NP effective couplings, which is -however- absent in the global fit results (\ref{global}). It must be noted that, being the $g-2$ kernel saturated at low energies, the observable in eq.~(\ref{eqamu}) is mostly sensitive to NP effects at low $\pi\pi$ invariant masses, which is precisely where the theory input is more reliable. Alternatively, one could think of comparing the energy dependence of both spectral functions directly, or use different kernels for enhancing the importance of the diverse energy regions but this would enlarge the incompatibility between different sets of $e^+e^-$ data that has been discussed largely during this workshop.\\

With respect to \textbf{b)}, the dispersion relation complies with analyticity and unitarity requirements, with its low-energy behaviour determined by chiral symmetry and its asymptotic limit by perturbative QCD. Whenever possible, experimental and lattice data are used as inputs to this framework \footnote{See talks by Gilberto Colangelo and Jos\'e Ram\'on Pel\'aez on applications of this formalism. In fact, the dispersive scalar and vector form factors that we employ in sections \ref{sec-pipi}, \ref{sec-etapi} and \ref{sec-Kpi} were extensively discussed in the talks by Sergi Gonz\`alez-Sol\'is and Emilie Passemar during this session.}. Specifically, the following form factors are used: the di-pion vector \cite{Dumm:2013zh}, the corresponding scalar \cite{Descotes-Genon:2014tla}, and the tensor form factor as derived in Ref.~\cite{Cirigliano:2017tqn} (normalized according to Ref.~\cite{Baum-Lubicz} at the origin). In this way, a fit to the invariant mass distribution and branching ratio measured by Belle \cite{Fujikawa:2008ma} yields \cite{Miranda:2018cpf}
\begin{equation} \label{result_eT}
 \hat{\epsilon}_T\,=\,(-1.3^{+1.5}_{-2.2})\times10^{-3}\,,
\end{equation}
provided $\hat{\epsilon}_S$ is restricted to realistic values ($|\epsilon_S|\lesssim 10^{-2}$ \cite{Bhattacharya_Cirigliano})~\footnote{The best fit result for $\hat{\epsilon}_T$ is confirmed using the reference results in Ref. \cite{SolisRoigpipiKK}.}.\\

Both approaches are challenged by the needed good control of the related systematic uncertainties but a conservative error estimation is done in both \textbf{a} and \textbf{b}.
\subsection{$\tau^-\to\eta\pi^-\nu_\tau$}
\label{sec-etapi}
Since the tensor form factor is basically proportional to the vector form factor (which usually dominates the dynamics of di-meson tau decays), the constraints on $\epsilon_T$ using tau data are quite competitive. On the contrary, it is not possible (in general) to obtain strong constraints on $\epsilon_S$. An exception, precisely, are the $\tau^-\to\eta\pi^-\nu_\tau$ decays \cite{Garces:2017jpz}, where the vector form factor contribution is suppressed by G-parity and the scalar form factor is unsuppressed by the di-meson squared mass difference, which is not a small parameter in this case. The corresponding hadronic input comes from the following form factors: \cite{Dumm:2013zh} vector, \cite{Escribano:2016ntp} scalar and tensor \cite{Baum-Lubicz, Cata:2007ns}. Only the upper limit obtained by BaBar \cite{delAmoSanchez:2010pc} allows us to set a very competitive limit \footnote{In the talk by Petar Rados this was shown for Belle-II data projections \cite{HernandezVillanueva:2018wbc}.} $\hat{\epsilon}_S\in[-0.83,0.37]\times10^{-2}$ ($\epsilon_S=(-6\pm15)\times10^{-2}$ according to Ref.~\cite{Cirigliano:2018dyk}, using our hadronic input).
\subsection{Summary of $|\Delta S|=0$ processes and comparison to other probes}
\label{sec-DeltaS0}
The combination of inclusive and exclusive tau decays analyzed in Ref.~\cite{Cirigliano:2018dyk} yields (in $10^{-2}$ units)
\begin{eqnarray}\label{global}
&& \epsilon_L^\tau-\epsilon_L^e+\epsilon_R^\tau-\epsilon_R^e\,=\,1.0\pm1.1\,,\quad \epsilon_R^\tau=0.2\pm1.3 \,,\nonumber\\
&& \epsilon_S^\tau=-0.6\pm1.5\,,\quad \epsilon_P^\tau=0.5\pm1.2\,,\quad \epsilon_T^\tau=-0.04\pm0.46\,, 
\end{eqnarray}
with correlations given in Ref.~\cite{Cirigliano:2018dyk}.
Complementary, our limits on the non-standard scalar and tensor interactions, obtained using strategy \textbf{b)} above are (with negligible correlations)
\begin{equation}
\hat{\epsilon}_S\in[-0.83,0.37]\times10^{-2}\; (\mathrm{at}\,68\%\, \mathrm{C.L.}) \quad\,,\quad \hat{\epsilon}_T\,=\,(-1.3^{+1.5}_{-2.2})\times10^{-3}\,.
\end{equation}

When the previous bounds (which are at the (sub)percent level) are interpreted in terms of anomalous spin-one couplings to the W in presence of BSM, a $\sim 2\sigma$ deviation is hinted \cite{Cirigliano:2018dyk}. The addition of hadronic $\tau$ decays to LHC and EWPO data improves the limits on the chirality of the W coupling to the lepton current with tau flavor \cite{Cirigliano:2018dyk}. These are -again- consistent with the SM at $\sim2\sigma$. Scales as high as $5,6$ TeV are currently probed in these analyses.

Let us finally emphasize that the limit on $\epsilon_S$ has $\sim 4$ times larger uncertainty than those coming from $\beta$ decays (including $0^+\to 0^+$ transitions, which do provide the best bounds, see an updated account in Ref.~\cite{Gonzalez-Alonso:2018omy}). The Belle-II measurement of the $\tau^-\to\eta\pi^-\nu_\tau$ decays together with an improved understanding of the dominant scalar form factor could make these decays the most precise scenario to restrict non-SM scalar interactions. For $\epsilon_T$, the uncertainty determined in $\tau^-\to\pi^-\pi^0\nu_\tau$ decays is approximately triple than the one in $\pi\to e\nu_e\gamma$ but di-pion tau decays can become the most precise determination if its spectrum is measured with great accuracy at Belle-II.
\subsection{$\tau^-\to K^-\nu_\tau$ decays}
\label{sec-K}
The material presented in this subsection has not appeared before in the literature, to my knowledge.

Analogously to $\tau^-\to\pi^-\nu_\tau$, $\tau^-\to K^-\nu_\tau$ restricts the combination $\epsilon_L^\tau-\epsilon_L^e-\epsilon_R^\tau-\epsilon_R^e-\frac{m_K^2}{M_\tau(m_u+m_s)}\epsilon_P^\tau$ \footnote{In the chiral limit, the coefficient of $\epsilon_P^\tau$ is the same as in eq.~(\ref{eq1pi}).}, with the $\epsilon$ coefficients corresponding to $u\to d$ transitions in section \ref{sec-pi} and to $u\to s$ transitions in this section (and the following ones).

Since the branching ratio for the one-Kaon tau decays has a triple relative error than the one for the one-pion ones, it may seem this cannot be a competitive source for extracting the previous combination of couplings. However, the error in eq.~(\ref{eq1pi}) is dominated by the lattice QCD determination of $F_\pi$, and $F_K$ is determined with a relative error $2.4$ times smaller than $F_\pi$ in the simulations \cite{Tanabashi:2018oca}, which partly compensates the former one. In the end, the abovementioned combination of couplings can be determined with an error $\sim 8\times 10^{-3}$ with its negative sign marginally favored, similarly to eq. (\ref{eq1pi}).

It is more interesting to look at the ratio between the branching fractions for $\tau^-\to K^-\nu_\tau$ and $K^-\to \mu^- \bar{\nu}_\mu$ decays, which yields
\begin{equation}\label{eqratiowithK}
 \epsilon_L^\tau-\epsilon_L^\mu-\epsilon_R^\tau+\epsilon_R^\mu-\frac{m_K^2}{M_\tau(m_u+m_s)}\epsilon^\tau_P+\frac{m_K^2}{m_\mu(m_u+m_s)}\epsilon^\mu_P\,=\,(-1.4\pm0.8)\times 10^{-2}\,,
\end{equation}
which is again dominated by the error on the branching fraction of the $\tau^-\to K^-\nu_\tau$ decays.

\subsection{$\tau^-\to (K\pi)^-\nu_\tau$ decays}
\label{sec-Kpi}
Strategy \textbf{b} above (section \ref{sec-pipi}) has just been employed for the $\tau^-\to(K\pi)^-\nu_\tau$ decays \cite{Rendon:2019awg}, where the hadronic form factors are obtained from Refs. \cite{Boito:2008fq} (vector) \footnote{We do not use either constraints from Kaon decays \cite{Boito:2010me} or from $\tau^-\to K^-\eta\nu_\tau$ decays \cite{Escribano:2014joa}.} \cite{Jamin:2006tj} (scalar) \footnote{This is fully determined from meson-meson scattering in the S-wave, as opposed to Ref. \cite{Antonelli-Cirigliano}.} and \cite{Baum-Lubicz, Cata:2007ns} (tensor). A fit to Belle \cite{Epifanov:2007rf} branching ratio and $K_S\pi^-$ spectrum yielded
\begin{equation}
 \hat{\epsilon}_S\,=\,(1.3\pm0.9)\times10^{-2}\,,\,\hat{\epsilon}_T\,=\,(0.7\pm1.0)\times10^{-2}\,,
\end{equation}
and made clear the $i=5,6,7$ Belle data points could not be explained either by heavy NP. Finally, we revisited the analysis of $A_{CP}$ in Ref. \cite{Cirigliano-CP} obtaining only a slightly larger upper limit for the NP contribution than in this reference. According to us $A_{CP}^{BSM}<10^{-6}$ \footnote{This agrees with the early bound obtained in Ref. \cite{Delepine:2006fv}.}, so it is impossible that heavy NP explains the $2.8\sigma$ discrepancy between the BaBar measurement, $A_{CP}=(-3.6\pm2.5)\times10^{-3}$ \cite{BABAR:2011aa}, and the SM prediction, $A_{CP}=(3.6\pm0.1)\times10^{-3}$ \cite{Grossman:2011zk} (see also \cite{Bigi:2005ts, Calderon:2007rg}) \footnote{Ref.~\cite{Devi:2013gya} proposed that tensor interactions could explain it, but Ref. \cite{Cirigliano-CP} show it was impossible, contrary to the recent claim in Ref. \cite{Dighe:2019odu}. In this case it should be noted -among other issues- that, in order to explain $A_{CP}^{exp}$ the size needed for the coefficient of the dimension $8$ operator is: on the one hand too large (it would break the power counting of the EFT) and, on the other, in conflict with other high- and low-energy data where it would contribute sizably.}. 
Inclusive analyses of the strange spectral function have not been pursued yet.
\subsection{Summary of $|\Delta S|=1$ processes and comparison to other probes}
\label{sec-DeltaS1}
It is very hard that any other probe competes with the extremely powerful limits obtained from Kaon decays \cite{Gonzalez-Alonso:2016etj} (as it is the case for hyperon decays \cite{Chang:2014iba}). Indeed, $K_{e3}$ decays (together with $0^+\to0^+$ processes) can probe physics at the amazing scale of $\mathcal{O}(500)$ TeV \cite{Gonzalez-Alonso:2018omy}. In the $\epsilon_S-\epsilon_T$ plane, $K_{\mu3}$ decays limit $\epsilon_S$ at levels beyond (direct or indirect) LHC reach \cite{Gonzalez-Alonso:2018omy}. However, for $\epsilon_T$, the allowed interval according to $\tau^-\to K_S\pi^-\nu_\tau$ decays only approximately doubles the one determined by LHC data, so a precision measurement of $K\pi$ tau decays at Belle-II can improve current constraints on this type of NP.
\section{Conclusions}
In this contribution I have reviewed why hadron tau decays are not only a clean QCD lab but also powerful NP probes by recalling what can be learnt using the low-energy limit of SMEFT for them. In Cabibbo favored processes they complement very nicely the information coming from EWPO and LHC data. Noteworthy, the upper limit on the $\tau^-\to\eta\pi^-\nu_\tau$ decays is already able to yield a very promising constraint for $\epsilon_S$. The discovery of this channel at Belle-II together with its improved theoretical understanding would allow for limits on $\epsilon_S$ at the level of the most accurate ones to date, from (nuclear) $\beta$ decays. Also, a more precise measurement of the di-pion $\tau$ decay mode at Belle-II would render the corresponding $\epsilon_T$ limits at the level of the currently most stringent ones, coming from radiative pion decays. For the strangeness-changing decays it is impossible to compete with the limits on $\epsilon_S$ coming from $K_{\ell3}$ decays but an improved measurement at Belle-II of the $\tau^-\to (K\pi)^-\nu_\tau$ decays spectrum may well imply a limit on $\epsilon_T$ at the level of $K_{\mu3}$ and LHC data. Although the vast majority of the results discussed in this talk appeared too late to be included in the Belle-II Physics book \cite{Kou:2018nap}, the rich potential of semileptonic tau decays as NP probes highlighted in this contribution should motivate their consideration as golden channels (in addition to $A_{CP}$ in the $K_S\pi$ decay modes) in the forthcoming years' searches.
\section*{Acknowledgements}
It is my pleasure to thank the BINP crew for organizing -once again- an excellent workshop and for making us feel at home in the late (and beautifully snowy) Siberian winter. I acknowledge my collaborators in the works presented in this talk. I also thank an enlightening exchange with Martin Hoferichter concerning the CP anomaly and possible BSM explanations. I benefited from suggestions by Gabriel L\'opez Castro on the draft of this contribution. Support from Conacyt projects 236394 and 250628 ('Ciencia B\'asica') and from Fondos Conacyt-SEP 2018 is acknowledged.

\end{document}